\documentclass{pasj00}
\begin{document}
\SetRunningHead{J.\ Fukue}
{Radiative Flow in a Luminous Disk}
\Received{2005/05/28}
\Accepted{2005/08/03}

\title{Radiative Flow in a Luminous Disk}

\author{Jun \textsc{Fukue}} 
\affil{Astronomical Institute, Osaka Kyoiku University, 
Asahigaoka, Kashiwara, Osaka 582-8582}
\email{fukue@cc.osaka-kyoiku.ac.jp}



\maketitle


\begin{abstract}
Radiatively-driven flow in a luminous disk
is examined in the subrelativistic regime of $(v/c)^1$,
taking account of radiation transfer.
The flow is assumed to be vertical, and
the gravity and gas pressure are ignored.
When internal heating is dropped,
for a given optical depth and radiation pressure
at the flow base (disk ``inside''), where the flow speed is zero,
the flow is analytically solved
under the appropriate boundary condition
at the flow top (disk ``surface''), where the optical depth is zero.
The loaded mass and terminal speed of the flow
are both determined by the initial conditions;
the mass-loss rate increases as the initial radiation pressure increases,
while the flow terminal speed increases
as the initial radiation pressure and the loaded mass decrease.
In particular, when heating is ignored,
the radiative flux $F$ is constant, and
the radiation pressure $P_0$
at the flow base with optical depth $\tau_0$ is bound 
in the range of $2/3 < cP_0/F < 2/3 + \tau_0$.
In this case,
in the limit of $cP_0/F = 2/3 + \tau_0$,
the loaded mass diverges and the flow terminal speed becomes zero,
while, in the limit of $cP_0/F = 2/3$,
the loaded mass becomes zero and the terminal speed approaches $(3/8)c$,
which is the terminal speed above the luminous flat disk
under an approximation of the order of $(v/c)^1$.
We also examine the case where
heating exists, and find that
the flow properties are qualitatively similar to the case
without heating.
\end{abstract}

\section{Introduction}

Accretion disks are tremendous energy sources in the active universe:
in young stellar objects, in cataclysmic variables,
in galaxtic X-ray sources and microquasars, and
in active galaxies and quasars
(see Kato et al. 1998 for a review).
In particular, in a supercritical accretion disk,
the mass-accretion rate highly exceeds the critical rate, 
the disk local luminosity exceeds the Eddington one,
and the mass loss from the disk surface
driven by radiation pressure takes place.

Such a radiatively driven outflow from a luminous accretion disk
has been extensively studied in the context of
models for astrophysical jets by many researchers
(Bisnovatyi-Kogan, Blinnikov 1977; Katz 1980; Icke 1980; Melia, K\"onigl 1989; 
Misra, Melia 1993; Tajima, Fukue 1996, 1998; Watarai, Fukue 1999;
Hirai, Fukue 2001; Fukue et al. 2001; Orihara, Fukue 2003),
as on-axis jets (Icke 1989; Sikora et al. 1996; Renaud, Henri 1998;
Luo, Protheroe 1999; Fukue 2005a),
as outflows confined by a gaseous torus
(Lynden-Bell 1978; Davidson, McCray 1980; Sikora, Wilson 1981; Fukue 1982),
or as jets confined by the outer flow or corona
(Sol et al. 1989; Marcowith et al. 1995; Fukue 1999),
and as numerical simulations
(Eggum et al. 1985, 1988).
In almost all of these studies, however,
the disk radiation field was treated as an external field,
and the radiation transfer was not solved.

The radiation transfer in the disk, on the other hand,
has been investigated in relation to the structure
of a static disk atmosphere and 
the spectral energy density from the disk surface
(e.g., Meyer, Meyer-Hofmeister 1982; Cannizzo, Wheeler 1984;
Shaviv, Wehrse 1986; Adam et al. 1988;
Hubeny 1990; Ross et al. 1992; Artemova et al. 1996;
Hubeny, Hubeny 1997, 1998; Hubeny et al. 2000, 2001;
Davis et al. 2005; Hui et al. 2005;
see also Mineshige, Wood 1990).
In these studies, however,
the vertical movement and mass loss were not considered.
Moreover, the relativistic effect for radiation transfer
was not considered in the current studies.

In this paper
we first examine the radiatively driven vertical flow
-- {\it moving photosphere} -- in a luminous flat disk
within the framework of radiation transfer
in the subrelativistic regime of $(v/c)^1$,
where we incorporate the effect of radiation drag
up to the first order of the flow velocity.
At the first stage,
we ignore the gravity of the central object as well as
the gas pressure, 
although internal heating is considered
in a limited way.
We then consider analytical solutions for such a radiative flow,
and examine the flow properties.

In the next section
we describe basic equations in the vertical direction.
In section 3
we examine the radiative flow without internal heating,
while the radiative flow with internal heating
is discussed in section 4.
The final section is devoted to concluding remarks.


\section{Basic Equations}

Let us suppose a luminous flat disk,
deep inside which
gravitational or nuclear energy is released
via viscous heating or other processes.
The radiation energy is transported in the vertical direction,
and the disk gas, itself, also moves in the vertical direction
due to the action of radiation pressure (i.e., plane-parallel approximation).
For simplicity, in the present paper,
the radiation field is considered to be sufficiently intense that
both the gravitational field of, e.g., the central object
and the gas pressure can be ignored:
tenuous cold normal plasmas in the super-Eddington disk,
cold pair plasmas in the sub-Eddington disk, or
dusty plasmas in the sub-Eddington disk.
As for the order of the flow velocity $v$,
we consider the subrelativistic regime,
where the terms of the first order of $(v/c)$ are retained,
in order to take account of radiation drag.
Under these assumptions,
the radiation hydrodynamic equations
for steady vertical ($z$) flows are described as follows
(Kato et al. 1998).

The continuity equation is
\begin{equation}
   \rho v = J ~(={\rm const.}),
\label{rho1}
\end{equation}
where $\rho$ is the gas density, $v$ the vertical velocity, and
$J$ the mass-loss rate per unit area.
The equation of motion is
\begin{equation}
   v\frac{dv}{dz} = \frac{\kappa_{\rm abs}+\kappa_{\rm sca}}{c}
                    \left[ F - (E+P)v \right],
\label{v1}
\end{equation}
where $\kappa_{\rm abs}$ and $\kappa_{\rm sca}$
are the absorption and scattering opacities (gray),
$E$ the radiation energy density, $F$ the radiative flux, and
$P$ the radiation pressure.
In a gas-pressureless approximation, the energy equation is reduced to
\begin{equation}
   0 = q^+ - \rho \left( j - c\kappa_{\rm abs} E
                  + \kappa_{\rm abs} \frac{2Fv}{c} \right),
\label{j1}
\end{equation}
where $q^+$ is the heating and $j$ is the emissivity.
For radiation fields, we have
\begin{equation}
   \frac{dF}{dz} = \rho
         \left[ j - c\kappa_{\rm abs} E
         + (\kappa_{\rm abs}-\kappa_{\rm sca}) \frac{Fv}{c} \right],
\label{F1}
\end{equation}
\begin{equation}
   \frac{dP}{dz} = \frac{\rho v}{c^2} \left( j - c\kappa_{\rm abs} E \right)
           -\rho \frac{\kappa_{\rm abs}+\kappa_{\rm sca}}{c}
                    \left[ F - (E+P)v \right],
\label{P1}
\end{equation}
and the closure relation,
\begin{equation}
   P = \frac{1}{3}E + \frac{4}{3}\frac{Fv}{c^2}.
\label{E1}
\end{equation}

Eliminating $j$ and $E$ with the help of equations (\ref{j1}) and (\ref{E1}),
and introducing the optical depth by
\begin{equation}
    d\tau = - ( \kappa_{\rm abs}+\kappa_{\rm sca} ) \rho dz,
\end{equation}
we can rearrange the basic equations up to the order of $(v/c)^1$ as
\begin{eqnarray}
   cJ\frac{dv}{d\tau} &=& -\left( F - 4Pv \right),
\label{v}
\\
   \frac{dF}{d\tau} &=& -\frac{ q^+ }
                              { (\kappa_{\rm abs}+\kappa_{\rm sca}) \rho }
                        +F\frac{v}{c},
\label{F2}
\\
   c\frac{dP}{d\tau} &=& F - 4Pv,
\label{P2}
\\
   (\kappa_{\rm abs}+\kappa_{\rm sca}) J \frac{dz}{d\tau} &=& -v.
\label{z}
\end{eqnarray}

Finally, 
integrating the sum of equations (\ref{v}) and (\ref{P2})
gives the momentum conservation in the present approximation,
\begin{equation}
   Jv + P = K ~(={\rm const.}),
\label{P}
\end{equation}
and similarly from equations (\ref{v}) and (\ref{F2})
we have the energy conservation,
\begin{equation}
   \frac{1}{2}Jv^2 + F = -\int \frac{ q^+ }
              { (\kappa_{\rm abs}+\kappa_{\rm sca}) \rho } d\tau,
\label{F}
\end{equation}
where the first term on the left-hand side is eventually dropped,
although we retain it here to clarify the physical meanings.

We solved equations (\ref{v}), (\ref{P}), and (\ref{F})
for appropriate boundary conditions, and
we obtained analytic solutions for several special cases.

As for the boundary conditions,
we imposed the following cases.
At the flow base (disk ``inside'')
with an arbitrary optical depth $\tau_0$,
the flow velocity is zero and
the radiation pressure is $P_0$,
where subscript 0 denotes the values at the flow base.
At the flow top (disk ``surface''),
where the optical depth is zero,
the radiation fields should satisfy the values
above the luminous infinite disk.
Namely, just above the disk with surface intensity $I_0$,
the radiation energy density $E_{\rm s}$, 
the radiative flux $F_{\rm s}$, and
the radiation pressure $P_{\rm s}$ are
$E_{\rm s}=(2/c)\pi I_0$,
$F_{\rm s}=\pi I_0$, and $P_{\rm s}=(2/3c)\pi I_0$, respectively,
where subscript s denotes the values at the flow top.
Rigorously speaking,
this latter boundary condition at the flow top
is for a static photosphere, and
should not be applied to a moving photosphere,
although we approximately used this condition 
in the present subrelativistic regime.
The exact boundary conditions are derived and discussed
in a separate paper for a fully special relativistic case.

As shown explicitly later,
in order for the flow to exist,
the radiation pressure $P_0$ at the flow base
should be restricted in some ranges.
The lower limit is set
due to the negative gradient of the radiation pressure
toward the flow top.
The upper limit is set
when the loaded mass diverges.
Beyond the upper limit,
the hydrostatic balance without mass loss
may be established in the vertical direction.

\section{Radiative Flow Without Heating}

We first examine the case,
where there is no viscous or nuclear heating: $q^+=0$.
In the luminous accretion disk,
it is usually supposed that
the heating processes take place deep inside the disk.
Hence, the assumption {\it without heating}
approximately describes the radiative flow
in or near the surface envelope of the disk.

In this case and under the present subrelativistic regime,
equation (\ref{F}) means that the radiative flux $F$
is conserved along the flow,
\begin{equation}
     F=F_{\rm s}.
\label{1F}
\end{equation}
Using the boundary conditions at the flow base ($v=0$),
equation (\ref{P}) is expressed as
\begin{equation}
     Jv+P=P_0.
\label{1P}
\end{equation}

Hence, equation (\ref{v}) becomes
\begin{equation}
   cJ\frac{dv}{d\tau} = -\left( F_{\rm s} - 4P_0 v \right).
\label{1v}
\end{equation}
This equation can be analytically solved to yield the solution for $v$,
\begin{equation}
   v = \frac{F_{\rm s}}{4P_0}
     \left[ 1 - e^{\frac{\displaystyle 4P_0}{\displaystyle cJ}
         (\displaystyle \tau - \tau_0)} \right].
\label{1v_solution}
\end{equation}
Thus, the radiative flow from the luminous disk without heating
is expressed in terms of the boundary values and the mass-loss rate.
In addition,
the flow velocity $v_{\rm s}$ at the flow top ($\tau=0$) is
\begin{equation}
   v_{\rm s} = \frac{F_{\rm s}}{4P_0}
     \left( 1 - e^{-\frac{\displaystyle 4P_0}{\displaystyle cJ}
         {\displaystyle \tau_0}} \right).
\label{1vs}
\end{equation}

Using the boundary condition at the flow top,
we further impose a condition on the values at boundaries.
As already mentioned, at the flow top
$cP=cP_{\rm s}=(2/3)F_{\rm s}$.
Hence, inserting boundary values into momentum equation (\ref{1P}),
we have the following relation:
\begin{equation}
    J \frac{F_{\rm s}}{4P_0}
     \left( 1 - e^{-\frac{\displaystyle 4P_0}{\displaystyle cJ}
         {\displaystyle \tau_0}} \right)
      + \frac{2}{3c}F_{\rm s} = P_0.
\label{1bc}
\end{equation}
That is, for given $\tau_0$ and $P_0$ at the flow base,
the mass-loss rate $J$ is determined in units of $F_{\rm s}/c^2$,
as an eigen value.

\begin{figure}
  \begin{center}
  \FigureFile(80mm,80mm){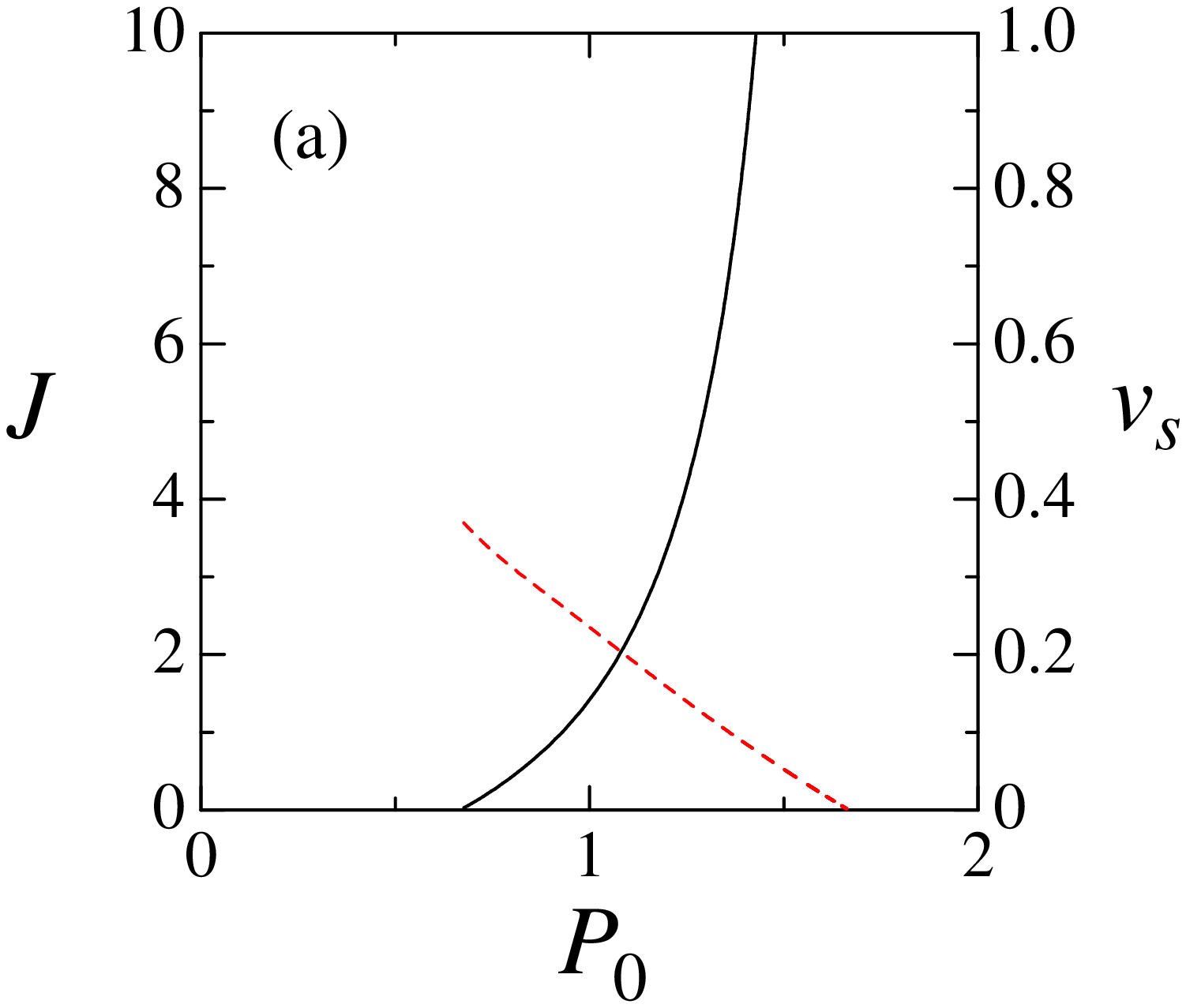}
  \end{center}
  \begin{center}
  \FigureFile(80mm,80mm){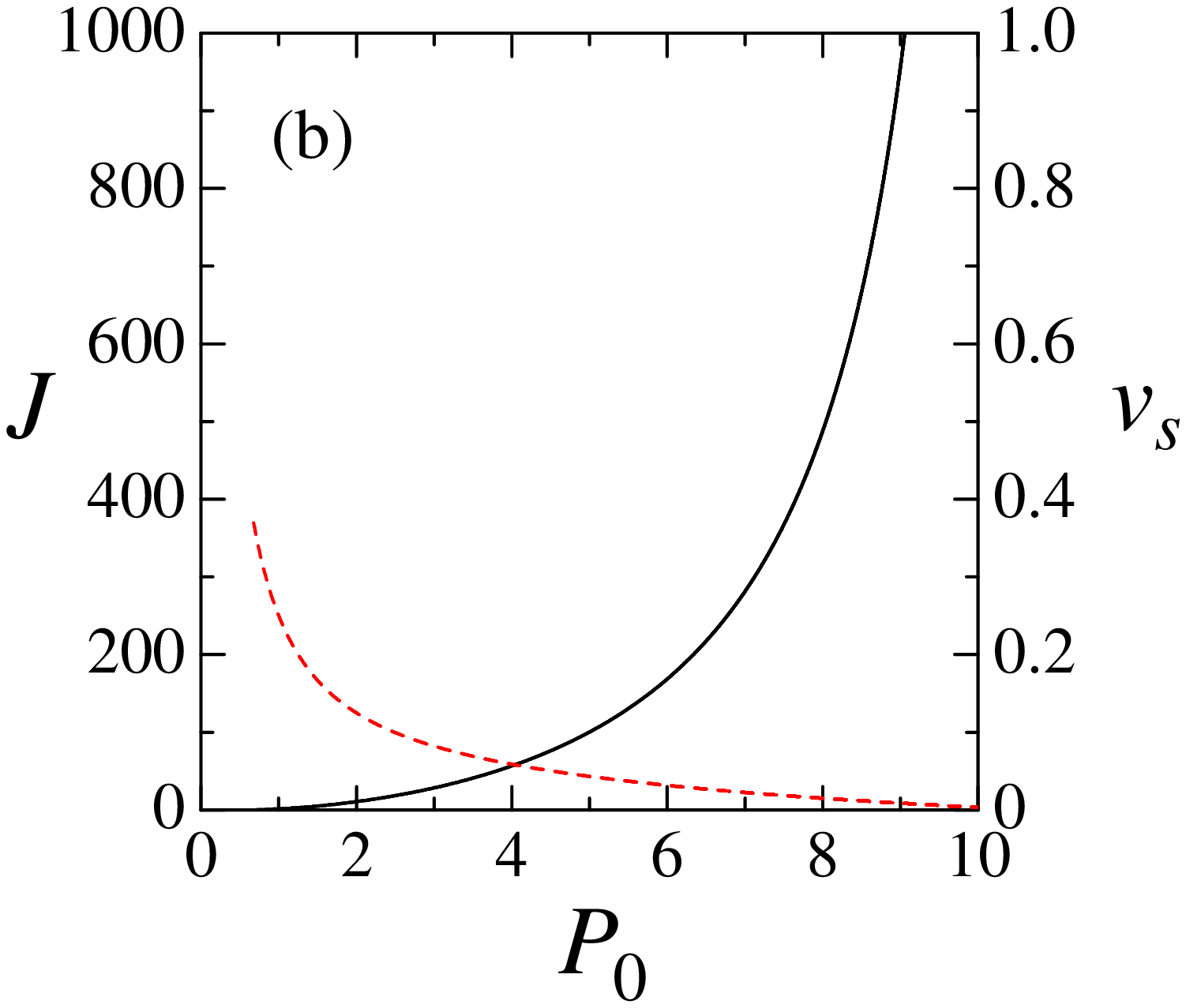}
  \end{center}
\caption{
Mass-loss rate $J$ (solid curves)
and terminal velocity $v_{\rm s}$ (dashed curves)
at the flow top as a function of $P_0$
for several values of $\tau_0$ at the flow base:
(a) $\tau_0=1$ and (b) $\tau_0=10$.
The quantities are normalized in units of $c$ and $F_{\rm s}$.
}
\end{figure}

In figure 1
we show the mass-loss rate $J$ (solid curves) 
and the terminal velocity $v_{\rm s}$ (dashed curves)
at the flow top as a function of $P_0$
for several values of $\tau_0$ at the flow base.
The quantities are normalized in units of $c$ and $F_{\rm s}$.

As can be seen in figure 1,
the mass-loss rate increases as the initial radiation pressure increases,
while the flow terminal speed increases
as the initial radiation pressure and the loaded mass decrease.

Moreover,
as can be seen in figure 1, and easily shown from equation (\ref{1bc}),
in order for the flow to exist,
the radiation pressure $P_0$ at the flow base is restricted
in some range,
\begin{equation}
   \frac{2}{3} < \frac{cP_0}{F_{\rm s}} < \frac{2}{3} +  \tau_0.
\end{equation}
At the upper limit of $cP_0/F_{\rm s} = 2/3 + \tau_0$,
the loaded mass diverges and the flow terminal speed becomes zero.
On the other hand, at the lower limit of $cP_0/F_{\rm s} = 2/3$,
the pressure gradient vanishes, 
the loaded mass becomes zero and the terminal speed approaches $(3/8)c$.
This is just the terminal speed of the gas,
driven by the radiation pressure under the radiation drag,
above the luminous flat infinite disk
in the present approximation of the order of $(v/c)^1$
(cf. Icke 1989).

\begin{figure}
  \begin{center}
  \FigureFile(80mm,80mm){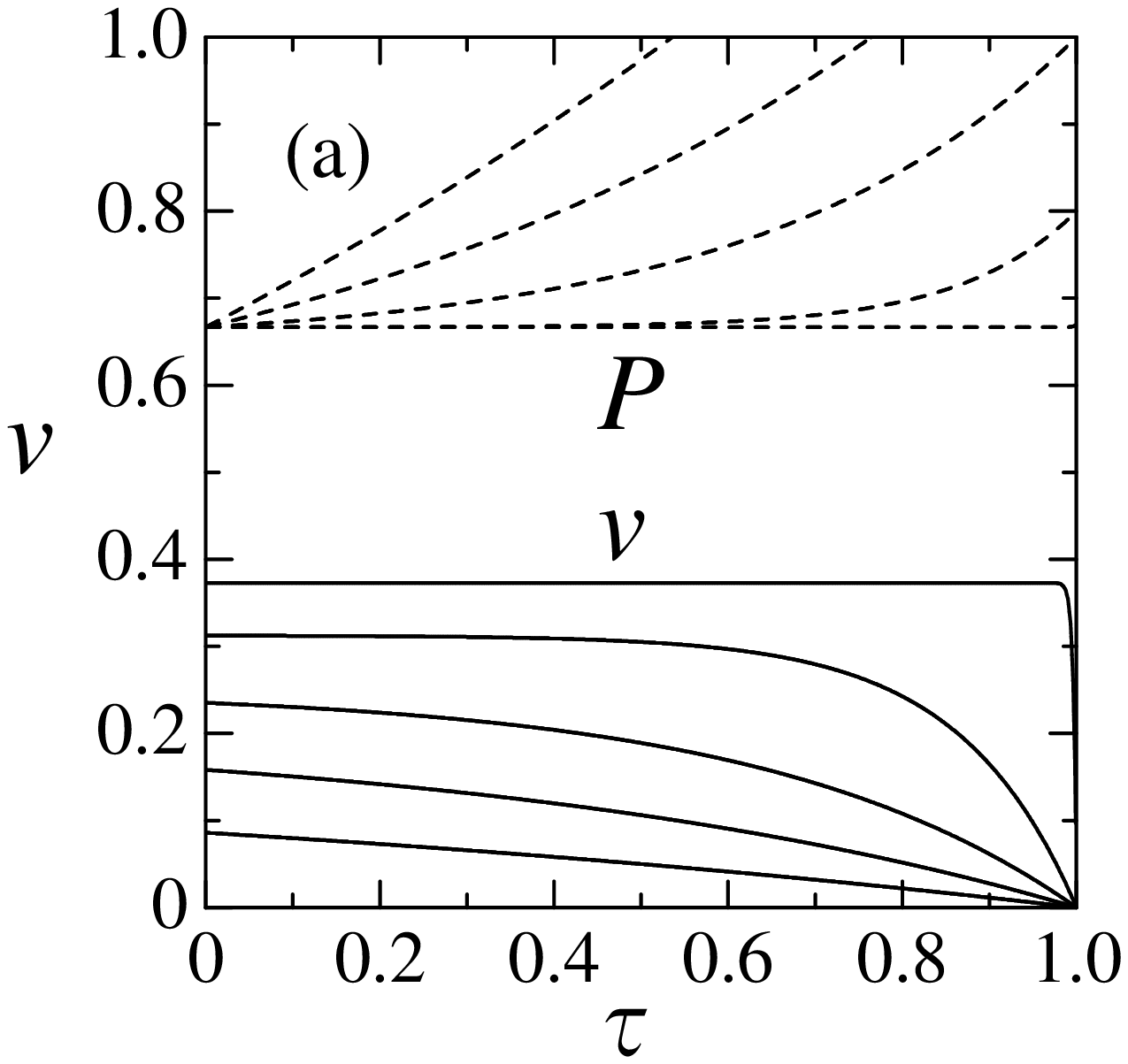}
  \end{center}
  \begin{center}
  \FigureFile(80mm,80mm){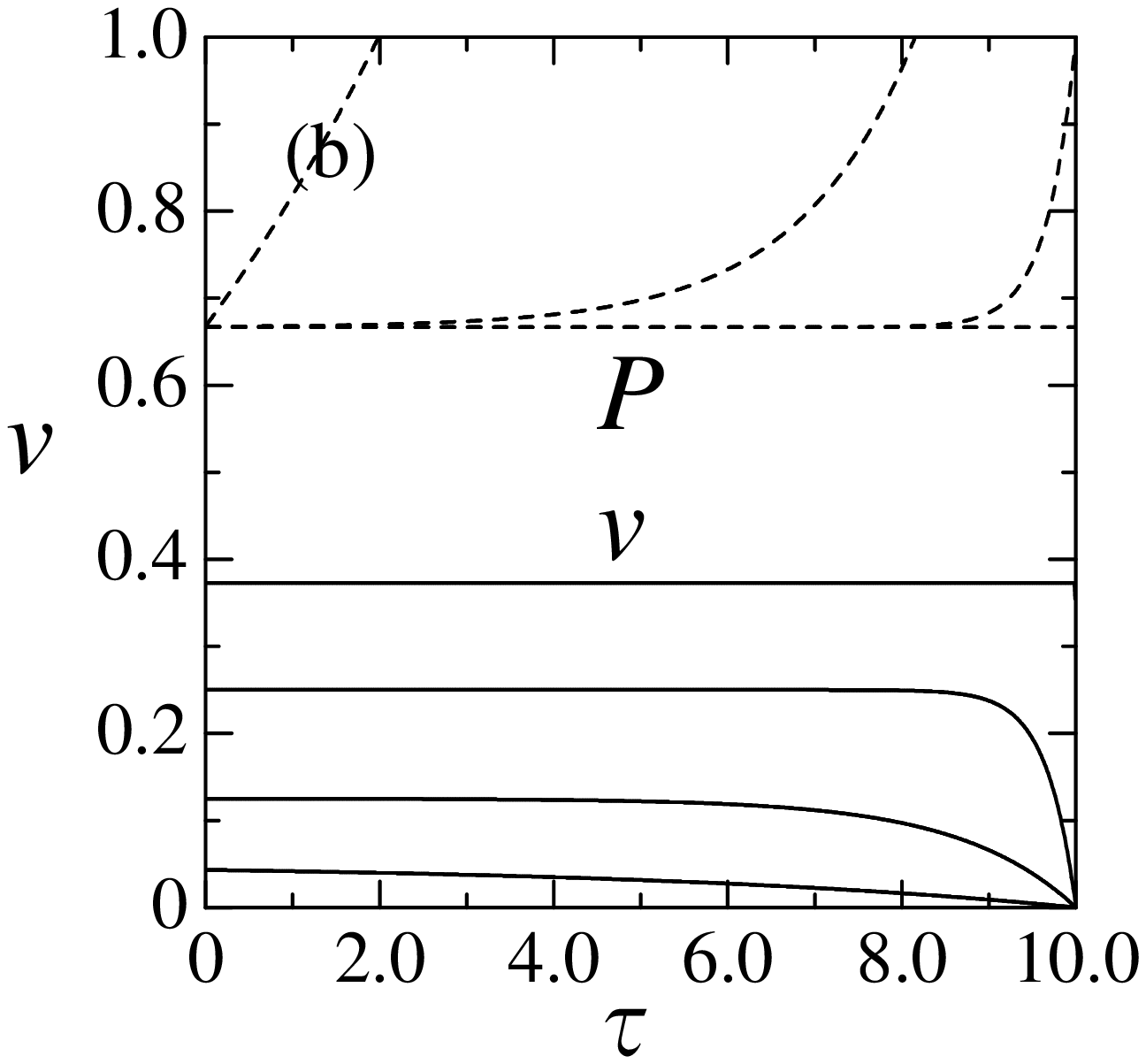}
  \end{center}
\caption{
Flow velocity $v$ (solid curves) 
and radiation pressure $P$ (dashed curves)
as a function of the optical depth $\tau$
for several values of $P_0$ at the flow base
in a few cases of $\tau_0$:
(a) $\tau_0=1$ and (b) $\tau_0=10$.
The quantities are normalized in units of $c$ and $F_{\rm s}$.
 From top to bottom of $v$ and from bottom to top of $P$,
the values of $P_0$ are
0.67, 0.8, 1, 1.2, 1.4 in (a), and
0.67, 1, 2, 5 in (b).
}
\end{figure}

In figure 2
we show the flow velocity $v$ (solid curves) 
and the radiation pressure $P$ (dashed curves)
as a function of the optical depth $\tau$
for several values of $P_0$ at the flow base
in a few cases of $\tau_0$.
The quantities are normalized in units of $c$ and $F_{\rm s}$.

When the initial radiation pressure $P_0$ at the flow base is large,
the pressure gradient between the flow base and the flow top is
also large.
As a result, the loaded mass $J$ also becomes large,
but the flow final speed $v_{\rm s}$ is small
due to momentum conservation (\ref{P}).
When the initial radiation pressure $P_0$ [$>(2/3c)F_{\rm s}$] is small,
on the other hand,
the pressure gradient becomes small, and
the loaded mass is also small, but
the flow final speed becomes large.
In the limiting case of $P_0=(2/3c)F_{\rm s}$,
the loaded mass $J$ becomes zero, but
the terminal speed $v_{\rm s}$ is $(3/8)c$,
as already mentioned.

Finally, inserting equation(\ref{1v_solution}) into equation (\ref{z}),
and integrating the resultant equation,
we obtain the height $z$ as a function of the optical depth $\tau$:
\begin{eqnarray}
  && \frac{(\kappa_{\rm abs}+\kappa_{\rm sca}) J}{c} z
\nonumber \\
  &=&
      \frac{F_{\rm s}}{4cP_0}
      \left\{ \left( \tau_0 - \tau \right)
              - \frac{cJ}{4P_0}
     \left[ 1 - e^{\frac{\displaystyle 4P_0}{\displaystyle cJ}
         (\displaystyle \tau - \tau_0)} \right]  \right\},
\end{eqnarray}
where we use the boundary condition at the flow base:
$z=0$ at $\tau=\tau_0$.
In addition,
the height $z_{\rm s}$ of the flow top (disk ``surface'') is also expressed as
\begin{equation}
   \frac{(\kappa_{\rm abs}+\kappa_{\rm sca}) J}{c} z_{\rm s} =
      \frac{F_{\rm s}}{4cP_0}
      \left( \tau_0 + \frac{2}{3} - \frac{cP_0}{F_{\rm s}} \right),
\end{equation}
where we use equation (\ref{1vs}).
Hence,
at the upper limit of $cP_0/F_{\rm s} = 2/3 + \tau_0$,
where the loaded mass diverges,
the height of the flow/disk becomes zero.
On the other hand, at the lower limit of $cP_0/F_{\rm s} = 2/3$,
where the loaded mass becomes zero,
the height of the flow/disk becomes infinite.

It should be noted that, at the upper limit of 
$cP_0/F_{\rm s} = 2/3 + \tau_0$ with a small terminal speed,
\begin{equation}
   T_{\rm c}^4 = \frac{3}{4} T_{\rm eff}^4
              \left( \frac{2}{3} + \tau_0 \right),
\end{equation}
where $T_{\rm c}$ is the temperature at the flow base
(i.e., $P_0=aT_{\rm c}^4/3$), and
$T_{\rm eff}$ the effective temperature at the flow top
(i.e., $F_{\rm s}=\sigma T_{\rm eff}^4$).
This is the Milne approximation.


\section{Radiative Flow With Heating}

We next examine the effect of heating: $q^+ \neq 0$.
That is, we consider the radiative flow
 from deep inside of the disk.

If the quantity $q^+/\rho$ is a function of $\tau$,
the heating term can be integrated
[see equations (\ref{F2}) or (\ref{F})].
In order for the integration to be performed analytically,
several functional forms need to be assumed.
One simple case, which we assume in the present paper,
is a following form:
\begin{equation}
   \frac{q^+}{ (\kappa_{\rm abs}+\kappa_{\rm sca}) \rho }
   = q_0 ~(={\rm const.}).
\end{equation}
This is quite reasonable since
the heating $q^+$ may be proportional to the density $\rho$.

In this case, under the present subrelativistic regime
and using the boundary conditions at the flow base,
equations (\ref{v}), (\ref{F}), and (\ref{P}) become, respectively,
\begin{eqnarray}
   cJ\frac{dv}{d\tau} &=& -\left( F - 4P_0 v \right),
\label{2v}
\\
   F &=& q_0 (\tau_0 - \tau) = 
         \frac{F_{\rm s}}{\tau_0} \left( \tau_0 - \tau \right),
\label{2F}
\\
   P &=& P_0 - Jv,
\label{2P}
\end{eqnarray}
where $F_{\rm s}$ ($=q_0 \tau_0$)
is again the radiative flux at the flow top (disk ``surface'').
Here, we impose the boundary condition on $F$ at the flow base
as $F=0$ at $\tau=\tau_0$.

The above equation (\ref{2v}) can be 
analytically solved to yield the solution for $v$:
\begin{equation}
   v = \frac{F_{\rm s}}{4P_0 \tau_0} \left( \tau_0 - \tau \right)
      - \frac{cJF_{\rm s}}{(4P_0)^2 \tau_0}
     \left[ 1 - e^{\frac{\displaystyle 4P_0}{\displaystyle cJ}
         (\displaystyle \tau - \tau_0)} \right].
\label{2v_solution}
\end{equation}
Thus, the radiative flow from the luminous disk with simple heating
is also expressed in terms of the boundary values and the mass-loss rate.
In addition,
the flow velocity $v_{\rm s}$ at the flow top ($\tau=0$) is
\begin{equation}
   v_{\rm s} = \frac{F_{\rm s}}{4P_0} 
      - \frac{cJF_{\rm s}}{(4P_0)^2 \tau_0}
     \left( 1 - e^{-\frac{\displaystyle 4P_0}{\displaystyle cJ}
         {\displaystyle \tau_0}} \right).
\label{2vs}
\end{equation}

Using the boundary condition at the flow top,
we further impose a condition on the values at the boundaries.
In this case with simple heating, we have the following relation:
\begin{equation}
    J \frac{F_{\rm s}}{4P_0}
      - \frac{cJ^2F_{\rm s}}{(4P_0)^2 \tau_0}
     \left( 1 - e^{-\frac{\displaystyle 4P_0}{\displaystyle cJ}
         {\displaystyle \tau_0}} \right)
      + \frac{2}{3c}F_{\rm s} = P_0.
\label{2bc}
\end{equation}
That is, for given $\tau_0$ and $P_0$ at the flow base,
the mass-loss rate $J$ is determined in units of $F_{\rm s}/c^2$.

\begin{figure}
  \begin{center}
  \FigureFile(80mm,80mm){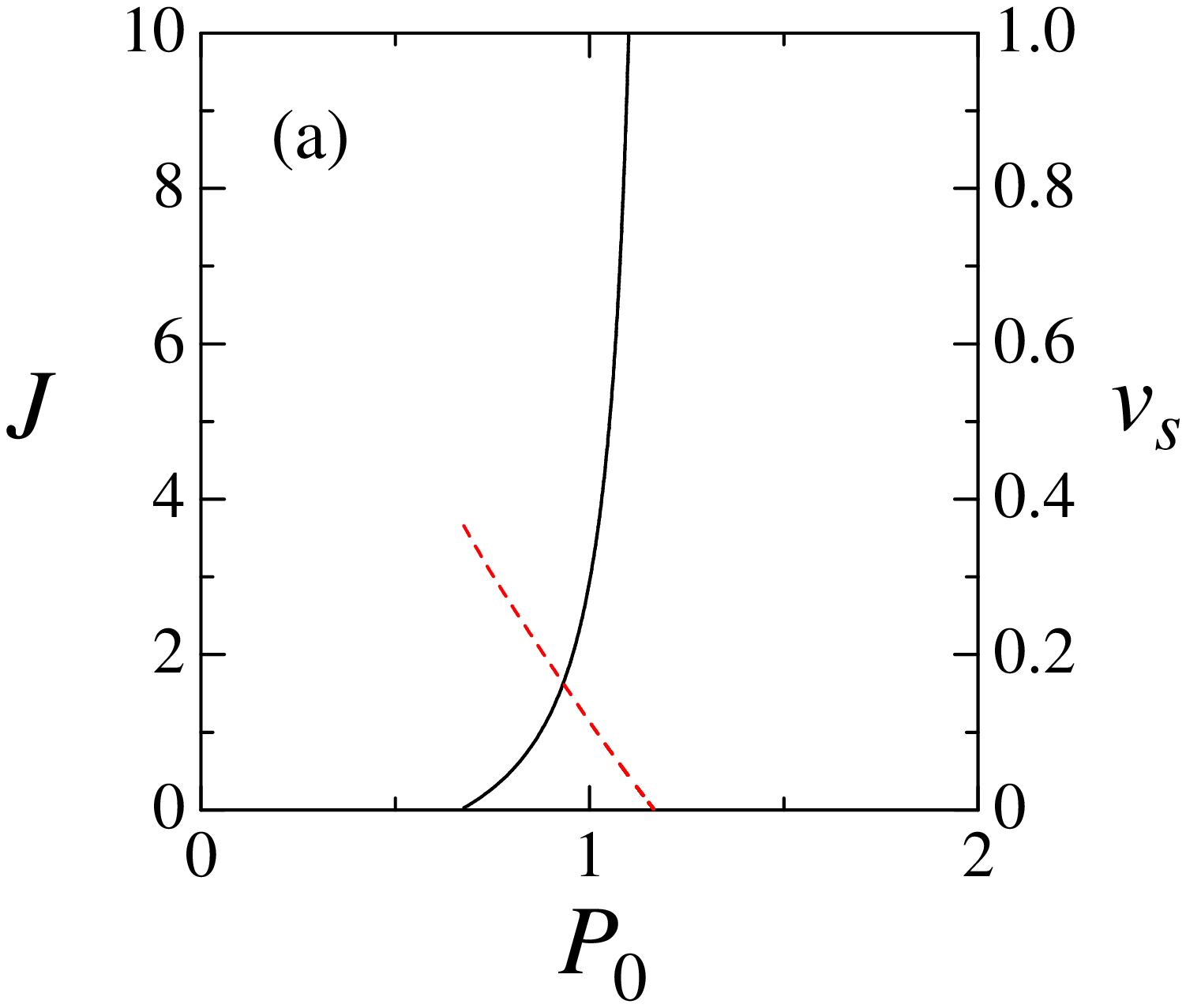}
  \end{center}
  \begin{center}
  \FigureFile(80mm,80mm){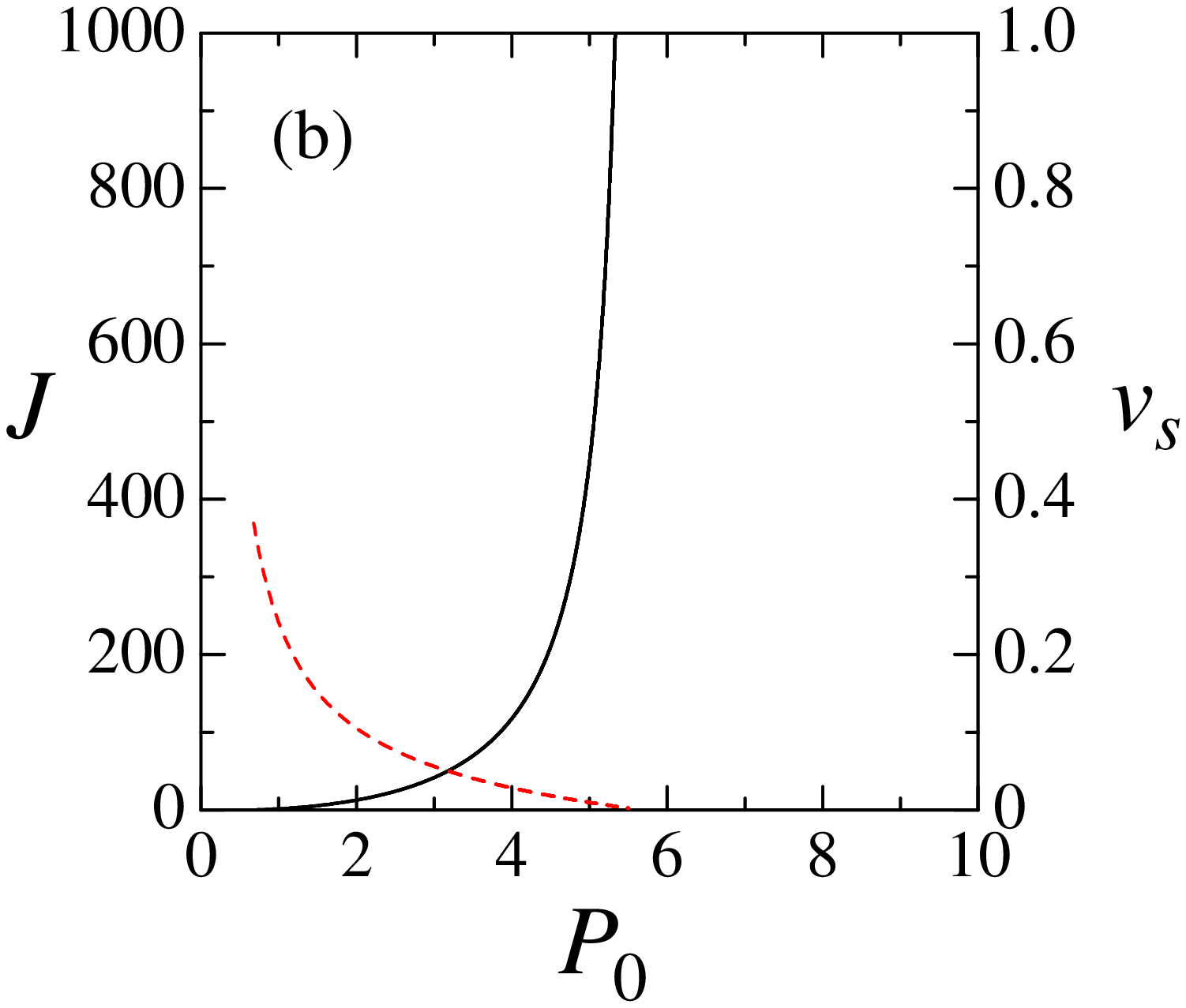}
  \end{center}
\caption{
Mass-loss rate $J$ (solid curves)
and terminal velocity $v_{\rm s}$ (dashed curves)
in the case with simple heating:
(a) $\tau_0=1$ and (b) $\tau_0=10$.
The quantities are normalized in units of $c$ and $F_{\rm s}$.
}
\end{figure}

In figure 3
we show the mass-loss rate $J$ (solid curves) 
and the terminal velocity $v_{\rm s}$ (dashed curves)
at the flow top as a function of $P_0$
for several values of $\tau_0$ at the flow base
in the case with simple heating.
The quantities are normalized in units of $c$ and $F_{\rm s}$.

The properties are qualitatively same as those without heating.
The mass-loss rate increases as the initial radiation pressure increases,
while the flow terminal speed increases
as the initial radiation pressure and the loaded mass decreases.
Moreover,
in order for flow to exist,
the radiation pressure $P_0$ at the flow base is restricted
in some range:
\begin{equation}
   \frac{2}{3} < \frac{cP_0}{F_{\rm s}} < \frac{2}{3} +  \frac{1}{2}\tau_0.
\end{equation}
At the upper limit,
the loaded mass diverges and the flow terminal speed becomes zero.
On the other hand, at the lower limit of $cP_0/F_{\rm s} = 2/3$,
the loaded mass becomes zero and the terminal speed approaches $(3/8)c$.

\begin{figure}
  \begin{center}
  \FigureFile(80mm,80mm){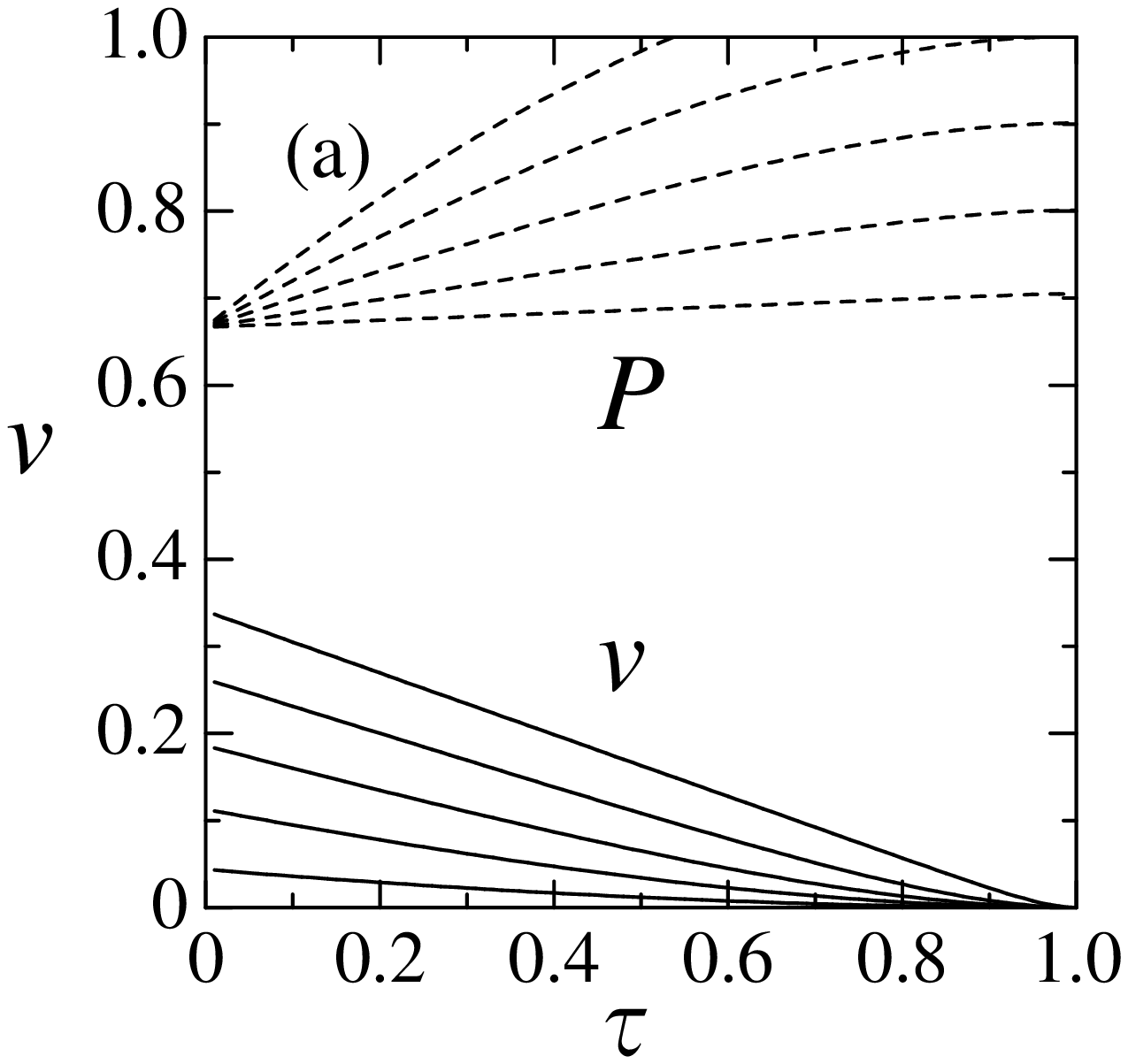}
  \end{center}
  \begin{center}
  \FigureFile(80mm,80mm){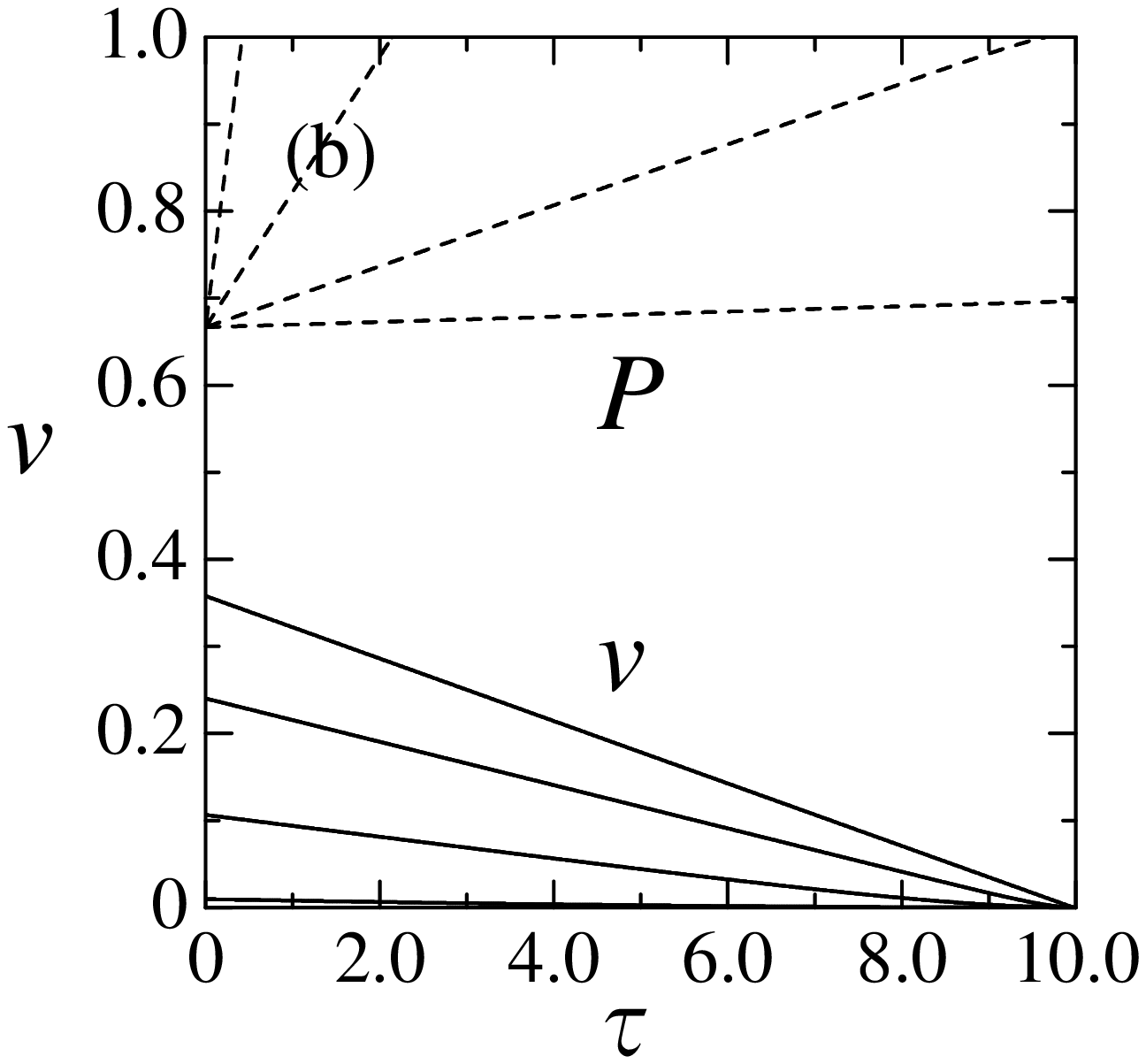}
  \end{center}
\caption{
Flow velocity $v$ (solid curves) 
and radiation pressure $P$ (dashed curves)
in the case with simple heating:
(a) $\tau_0=1$ and (b) $\tau_0=10$.
The quantities are normalized in units of $c$ and $F_{\rm s}$.
 From top to bottom of $v$ and from bottom to top of $P$,
the values of $P_0$ are
0.7, 0.8, 0.9, 1, 1.1 in (a), and
0.7, 1, 2, 5 in (b).
}
\end{figure}

In figure 4
we show the flow velocity $v$ (solid curves) 
and the radiation pressure $P$ (dashed curves)
as a function of the optical depth $\tau$
for several values of $P_0$ at the flow base
in a few cases of $\tau_0$
in the case with simple heating.
The quantities are normalized in units of $c$ and $F_{\rm s}$.

In this case with simple heating,
the velocity distribution is somewhat different from
that in the case without heating.
In the case without heating,
the flow is rather quickly accelerated to be a terminal speed,
since the radiative flux is constant.
In the case with heating, on the other hand,
the flow is gradually accelerated,
since the radiative flux is increasing from the flow base
toward the flow top due to heating.

Finally, inserting equation(\ref{1v_solution}) into equation (\ref{z}),
and integrating the resultant equation,
we obtain the height $z$ as a function of the optical depth $\tau$:
\begin{eqnarray}
  && \frac{(\kappa_{\rm abs}+\kappa_{\rm sca}) J}{c} z
\nonumber \\
  &=& 
      \frac{F_{\rm s}}{4cP_0}
      \left\{ \frac{(\tau-\tau_0)^2}{2\tau_0}
              +\frac{cJ}{4P_0 \tau_0} \left( \tau_0 - \tau \right)
      \right.
\nonumber \\
   && \left.
              + \left( \frac{cJ}{4P_0} \right)^2 \frac{1}{\tau_0}
     \left[ 1 - e^{\frac{\displaystyle 4P_0}{\displaystyle cJ}
         (\displaystyle \tau - \tau_0)} \right]  \right\},
\end{eqnarray}
where we use the boundary condition at the flow base:
$z=0$ at $\tau=\tau_0$.
In addition,
the height $z_{\rm s}$ of the flow top (disk ``surface'') is also expressed as
\begin{equation}
   \frac{(\kappa_{\rm abs}+\kappa_{\rm sca}) J}{c} z_{\rm s} =
      \frac{F_{\rm s}}{4cP_0}
      \left( \frac{\tau_0}{2} + \frac{2}{3} -\frac{cP_0}{F_{\rm s}} \right),
\end{equation}
where we use equation (\ref{2vs}).
Hence,
at the upper limit of $cP_0/F_{\rm s} = 2/3 + \tau_0$/2,
where the loaded mass diverges,
the height of the flow/disk becomes zero.
On the other hand, at the lower limit of $cP_0/F_{\rm s} = 2/3$,
where the loaded mass becomes zero,
the height of the flow/disk becomes infinite.

It should be noted that, at the upper limit of 
$cP_0/F_{\rm s} = 2/3 + \tau_0$/2 with small terminal speed,
\begin{equation}
   T_{\rm c}^4 = \frac{3}{4} T_{\rm eff}^4
              \left( \frac{2}{3} + \frac{1}{2}\tau_0 \right),
\end{equation}
where $T_{\rm c}$ is the temperature at the flow base
(i.e., $P_0=aT_{\rm c}^4/3$), and
$T_{\rm eff}$ the effective temperature at the flow top
(i.e., $F_{\rm s}=\sigma T_{\rm eff}^4$).
This is just the modified Milne approximation.

\section{Concluding Remarks}

In this paper 
we have examined the radiative flow from a luminous disk,
while taking account of the radiative transfer,
in the subrelativistic regime of $(v/c)^1$.
The flow velocity, the radiation pressure distribution, 
and other quantities
are analytically solved as a function of the optical depth
for the cases without and with heating.
At the flow base (disk ``inside''), where the flow speed is zero,
the initial optical depth is $\tau_0$ and
the initial radiation pressure is $P_0$;
in the usual accretion disk
these quantities are determined in terms of
the central mass, the mass-accretion rate, and the viscous process
as a function of the radius.
That is,
the optical depth is determined by the disk surface density,
the radiative flux is given by the central mass and the accretion rate,
and
the radiation pressure relates to the disk internal structure.
At the flow top (disk ``surface''), where the optical depth is zero,
the radiation fields are assumed to be those
above a static flat disk with uniform intensity.
In order to match this boundary condition,
one relation is imposed on the boundary quantities, and
the mass-loss rate $J$ and terminal speed $v_{\rm s}$
are both determined by the initial conditions $\tau_0$ and $P_0$,
as eigenvalues.
In particular, in order for the flow to exist,
the initial radiation pressure is bound in some range:
at the upper limit,
the loaded mass diverges and the flow terminal speed becomes zero,
while, at the lower limit,
the loaded mass becomes zero and the terminal speed approaches $(3/8)c$,
which is the terminal speed above the luminous disk
under the approximation of the order of $(v/c)^1$.

Among various types of astrophysical jets and outflows,
some are believed to be accelerated by continuum radiation,
and to which the present mechanism can be applied.
For example, in supersoft X-ray sources,
including accreting white dwarfs,
high-velocity mass outflows were reported
(e.g., Cowley et al. 1998).
The wind velocity measured in these objects is about
3000--5000~km~s$^{-1}$,
which is of the order of the escape velocity of white dwarfs.
In addition, the line profile depends on the inclination angle, and
the P~Cyg profiles are seen.
Hence, in supersoft X-ray sources,
the outflows should blow off from the innermost accretion disk.
However, the gas temperature of the inner disk is so high
that the line-driven mechanism cannot operate, and therefore,
mass outflows in these objects would be driven
by continuum radiation.

In SS~433, a prototype of astrophysical jets,
the mass-accretion rate highly exceeds the Eddington one,
and twin jets with $0.26~c$ would be accelerated by radiation pressure
(see, e.g., Cherepashchuk et al. 2005 for a recent obsevation).
The luminosity of microquasar GRS~1915$+$105
exceeds the Eddington one ($L_{\rm Edd}$) during its outburst.
In quiscent phase, its luminosity never drop
below $\sim 0.3 L_{\rm Edd}$
(Done, Gierli\'nski 2004).

In broad absorption-line quasars (BAL QSOs),
strong broad absorption lines in the UV resonance
are always blueshiftd relative to the emission-line rest frame,
which indicates the presence of outflows from the nucleus,
with velocities as large as $0.2~c$.
These moderately high-velocity outflows 
are supposed to be accelerated by line or continuum radiation
(e.g., Murray et al. 1995; Proga et al. 2000; Proga, Kallman 2004).
In these objects, the outflows must consist of dense clouds, or
be shielded from the central strong source,
in order to operate a line-driven mechanism.
Otherwise, the material is highly ionized by the central source,
and, instead of line-acceleration, continuum acceleration may work
(see, e.g., Chelouche, Netzer 2003).

Finally, narrow-line Seyfert 1 galaxies and luminous quasars
are also believed to harbor supercritically accreting black holes,
and high-velocity outflows are reported as warm absorbers
(cf. Blustin et al. 2005).
In these objects, radiative acceleration by continuum
may play a dominant role.

The radiative flow investigated in the present paper
must be quite {\it fundamental problems} for
accretion disk physics and astrophysical jet formation,
although the present paper is only the first step, and
there are many simplifications at the present stage.

For example, we have ignored the gravitational field
produced by the central object.
This means that the flow considered in the present paper
would correspond to normal plasmas in the super-Eddington disk,
pair plasmas in the sub-Eddington disk, or
dusty plasmas in the luminous disk.
In other cases, or even for normal plasmas in the super-Eddington disk,
the gravitational field would affect the flow properties.
In particular, the gravitational field in the vertical direction
is somewhat complicated (e.g., Fukue 2002, 2004),
and the influence of the gravitational field is important.
The effect of gravity will be considered
in a seperate paper (Fukue 2005b).

We also ignored the gas pressure.
This means that the radiation field is sufficiently intense.
In general cases, where the gas pressure is considered,
there usually appear sonic points (e.g., Fukue 2002, 2004),
and the flow is accelerated from subsonic to supersonic.
The cross section of the flow also has a similar influence.
In this paper we consider a purely vertical flow,
and the cross section of the flow is constant.
If the cross section of the flow increases along the flow,
the flow properties, such as the transonic nature, would be influenced.

In the present paper,
we have considered the internal heating processes in a simple form.
Since the heating prcesses should couple with the viscous process
or other heating processes, such as a nuclear process,
we should treat internal heating more carefully.

In order to take account of the effect of radiation drag,
we consider the subrelativistic regime,
where the terms of the first order of $(v/c)$ are retained.
As a result, the terminal speed $(3/8)c$ for the subrelativistic regime
appears in the extreme case.
This is quite consistent with the previous knowledge.
As is well known,
the terminal speed above the luminous flat disk in the full relativity
is $[(4-\sqrt{7})/3]c \sim 0.45c$ (Icke 1989).
To consider highly relativistic cases,
all of the terms up to $(v/c)^2$ should be retained
(cf. Kato et al. 1998).
A related problem is the boundary condition at the flow top.
In the present paper, we impose the boundary condition such that
the radiation fields at the flow top are those 
above the luminous flat disk.
Rigorously speaking, this boundary condition should be modified,
since the disk gas, itself, is now moving.
In the highly relativistic case, the deviation would be serious.
The exact boundary conditions are derived and discussed
in a separate paper for a fully special relativistic case
(Fukue 2005c).

\vspace*{1pc}

This work has been supported in part
by a Grant-in-Aid for the Scientific Research (15540235 J.F.) 
of the Ministry of Education, Culture, Sports, Science and Technology.


\end{document}